\documentclass[aps,prl,twocolumn,amsmath,amssymb]{revtex4}
\usepackage{graphicx}

\begin{document}

\date{\today}
\title{Mobility in semiconducting carbon nanotubes at finite carrier density}

\author{Vasili Perebeinos$^*$, J. Tersoff, and Phaedon Avouris}
\affiliation{IBM Research Division, T. J. Watson Research Center,
Yorktown Heights, New York 10598}

\date{\today}

\begin{abstract}
Carbon nanotube field-effect transistors operate over a wide range
of electron or hole density, controlled by the gate voltage. Here
we calculate the mobility in semiconducting nanotubes as a
function of carrier density and electric field, for different tube
diameters and temperature. The low-field mobility is a
non-monotonic function of carrier density, and varies by as much
as a factor of 4 at room temperature. At low density, with
increasing field the drift velocity reaches a maximum and then
exhibits negative differential mobility, due to the
non-parabolicity of the bandstructure. At a critical density
$\rho_c\sim$ 0.35-0.5 electrons/nm, the drift velocity saturates
at around one third of the Fermi velocity. Above $\rho_c$, the
velocity increases with field strength with no apparent
saturation.
\end{abstract}
\maketitle

Carbon nanotubes have enabled new nano-electronic devices such as
quasi-one-dimensional field-effect transistors (CN-FET)
\cite{Avouris} and electro-optical devices
\cite{Misewich,Marcus2}. Ballistic devices have generated
excitement as the ideal limit of transistor performance
\cite{Javey1}. Yet diffusive devices remain important
\cite{Marcus2,Snow,Grunner} due to the very high mobility even at
room temperature \cite{Durkop}. Much effort has gone into
determining their transport properties experimentally
\cite{Yao,Javey,Park,Zhou,ChenF} and theoretically
\cite{Pennington,Perebeinos3,Guo,Verma}.

Conventional transistors operate by control of carrier density,
i.e. band filling via an external gate electrode and its
capacitive coupling to the conduction channel. Yet while carrier
density is a crucial variable in device operation, relatively
little is known about how the mobility in nanotube transistors
varies with carrier density, especially at high fields. Only
recently have systematic studies of the dependence on gate voltage
been reported \cite{Zhou,ChenF}.

Here we calculate the mobility in carbon nanotubes as a function
of charge carrier density. We find that the mobility depends
sensitively on carrier density. Specifically, at low field and low
temperature, the mobility initially increases with carrier
density, because the phase space for scattering \cite{disclaimer}
is reduced by band filling. The mobility then decreases abruptly
as the band filling reaches the second band, because a new channel
for scattering opens up. The low-field mobility can be expressed
as the product of scattering time and hinverse effective mass,
where the effective mass is nearly independent of temperature, and
increases smoothly with band filling. Our results quantify a
recent analysis of low-field mobility \cite{Zhou}.

At high fields, we find negative differential mobility (NDM) at
low density, decreasing and disappearing with increasing density.
Surprisingly, this NDM appears to be unrelated to the occupation
of the second band, but can be ascribed to the nonparabolicity of
the first band. The minimum differential mobility, i.e. maximum
magnitude of the NDM, increases with diameter and decreases with
temperature, with a systematically larger value in tubes with
mod$(n-m,3)=1$ than in those with mod$(n-m,3)=-1$. This is due to
the stronger band nonparabolicity in the former.

Our calculations use a  tight-binding description for the
electronic structure \cite{Saito}. The phonons are described using
a model similar to that of Aizawa {\it et al.}~\cite{Aizawa}. We
model the electron-phonon  interaction by the Su-Schrieffer-Heeger
 model \cite{Su}, with matrix element $t=t_0-g\delta u$
dependent on the change of the nearest neighbor C-C distance
($\delta u$), where $t_0=3$ eV. We take the electron-phonon
coupling constant to be $g=5.3$ eV/\AA~\cite{Perebeinos2}.

We solve the steady-state multi-band Boltzmann equation in the
presence of an electric field, to determine the electron
distribution function $g_k$:
\begin{eqnarray}
\frac{eE}{\hbar }\frac{\partial g_{k}}{\partial k}=-\sum_{q}&&[
W_{k,k+q}g_{k}\left( 1-g_{k+q}\right)
\nonumber\\
&&-W_{k+q,k}g_{k+q}\left( 1-g_{k}\right)], \label{eq4}
\end{eqnarray}
where $W_{k,k+q}$ is an electron-phonon scattering rate
\cite{Perebeinos3}. The indices k and q label both the continuous
wave vector along the tube axis and the band index. Using the
non-equilibrium distribution function $g_k$ from Eq.~(\ref{eq4}),
we calculate the drift velocity $V_d$ and the mobility $\mu$
according to:
\begin{eqnarray}
V_d&=& \mu E= \frac{1}{g_t}\sum_k V_k g_k
 \label{eqmass}
\end{eqnarray}
where $V_k$ is the band velocity and $g_t=\sum_k g_k$ is the band
filling factor (electrons/carbon).

\begin{figure}[h!]
\includegraphics[height=2.8in]{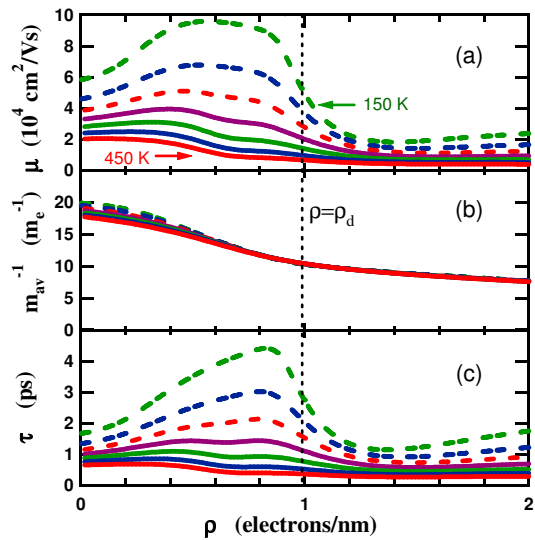}
\caption{\label{figdecom} (a) Low-field electron mobility in
(19,0) tube ($d$=1.5 nm) as a function of the charge carrier
density $\rho$. (b-c) Decomposition of mobility into a product of
(b) inverse effective mass and (c) scattering time. Temperatures
are (from top to bottom in each frame) T=150 K (dashed green),
T=200 K (dashed blue), T=250 K (dashed red), T=300 K (solid
magenta), T=350 K (solid green), T=400 K (solid blue), T=450 K
(solid red). The vertical dashed line is at $\rho=\rho_d=0.99$
electrons/nm (see text). The density of 1 electrons/nm correspond
to band filling of 0.0057 electrons/carbon for a (19,0) tube.}
\end{figure}

The low-field electron mobility is shown in Fig.~\ref{figdecom}a
for a (19,0) tube. (The hole mobility is identical, due to the
electron-hole symmetry in our model.) The mobility has a complex
and highly non-monotonic dependence on density at low temperature,
and decreases at higher temperature due to phonon scattering.

To understand the low-field mobility, we use the classical
relationship $\mu=e\tau/m_{av}$ to decompose the mobility into a
product of two terms: the scattering time $\tau$ and the averaged
inverse effective mass $m_{av}^{-1}$. The natural definition of
$m_{av}$ and resulting definition of $\tau$ are
\cite{disclaimer3}:
\begin{eqnarray}
\frac{1}{m_{av}}&=& \frac{1}{g_t}\sum_k g_k \frac{\partial^2
\varepsilon_k}{\partial (\hbar \kappa)^2} \ , \ \
\tau=\frac{m_{av}\mu}{e}. \label{eqmom}
\end{eqnarray}
The effective mass contribution to the low-field mobility is shown
in Fig.~\ref{figdecom}b - it is virtually independent of
temperature and it is a monotonically decreasing function of the
electron density $\rho$. In contrast, the scattering time
contribution in Fig.~\ref{figdecom}c has a strong dependence on
both temperature and charge density. The charge scale is
determined primarily by the density $\rho_d$ at which the first
band is filled up to the energy of the bottom of the second band
(at $T=0$ K). The scattering time drops sharply around $\rho_d$,
reaching a minimum before rising again.

At low field, $g_k$ is given essentially by the Fermi-Dirac
distribution, and we can solve Eq.~(\ref{eqmom}) for $m_{av}$
analytically for a given band dispersion $\varepsilon_k$. We use a
hyperbolic band dispersion
\begin{eqnarray}
\varepsilon_k=\sqrt{\Delta^2+\Delta\frac{\hbar^2 \kappa^2}{m_0}},
\label{eq1}
\end{eqnarray}
which provides a good approximation to the tight-binding bands in
semiconducting carbon nanotubes. Here $\Delta$ is half the bandgap
energy and $m_0$ is the effective mass at the bottom of the band.
At zero temperature, we solve Eq.~(\ref{eqmom}) and
Eq.~(\ref{eq1}) for the density-dependent effective mass:
\begin{eqnarray}
\frac{m_{av}(T=0)}{m_0}=\sqrt{1+\frac{\pi^2\hbar^2\rho^2}{16\Delta
m_0}}, \label{eqma}
\end{eqnarray}
where the relationship $\kappa_F=\pi\rho/4$ between the 1D Fermi
vector $\kappa_F$ and density $\rho$ has been used. As the Fermi
level increases with density, the effective mass gradually
increases and almost doubles when the Fermi level reaches the
bottom of the second band (see Fig.~\ref{figdecom}b).

With increasing density, the scattering time initially increases.
This can be understood as resulting from the decreasing density of
states available for scattering. With further density increase,
electrons start to populate the second band and an additional
channel opens up for scattering. This reduces the scattering time
at all temperatures, with the largest change occurring at low
temperature. As the temperature increases, the density dependence
of the scattering time becomes weaker. Therefore, above room
temperature, both the effective mass and the scattering time make
comparable contributions to the density dependence of the
low-field mobility. In contrast, at low temperatures, the
scattering time dominates the density dependence of the low-field
mobility.

The non-monotonic density dependence of the mobility, predicted
here, agrees well with experiment \cite{Durkop,Zhou}. Experiment
reports a maximum in the mobility and a few-fold drop from its
peak value as the gate voltage increases by several volts. This
behavior is nicely reproduced in Fig.~\ref{figdecom}. Agreement in
the absolute value of the mobility is not expected, because we
only consider phonon scattering, while additional scattering
mechanisms (e.g. impurities, charged defects in the oxide) are
present in the experiment.

\begin{figure}[h!]
\includegraphics[height=2.8in]{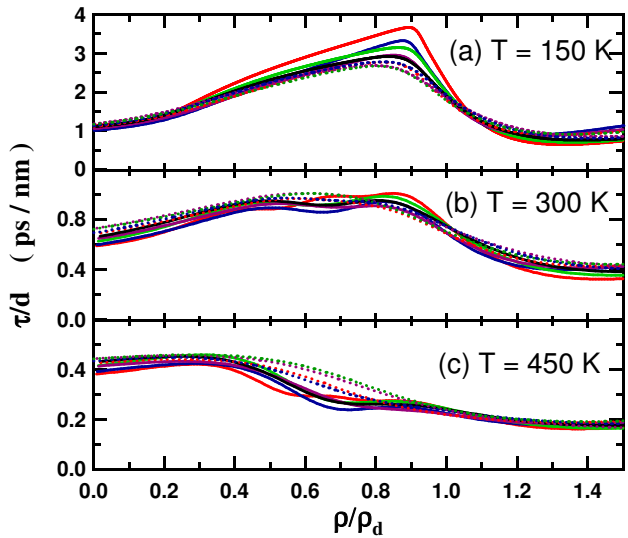}
\caption{\label{figtau} Diameter scaling of the scattering time in
carbon nanotubes. We show scaled scattering time $\tau/d$ versus
scaled density $\rho/\rho_d$ for zig-zag (n,0) tubes with diameter
range $d=$ 1.0 - 2.0 nm. In electrons/nm, $\rho_d=1.44$ for n=13
(solid red), $\rho_d=1.33$ for 14 (solid blue), $\rho_d=1.17$ for
16 (solid green), $\rho_d=1.10$ for 17 (solid magenta),
$\rho_d=0.99$ for 19 (solid black), $\rho_d=0.94$ for 20 (dashed
red), $\rho_d=0.85$ for 22 (dashed blue), $\rho_d=0.81$ for 23
(dashed green), $\rho_d=0.75$ for 25 (dashed magenta) at
temperatures (a) $T=150$ K, (b) $T=300$ K, (c) $T=450$ K.}
\end{figure}

Carbon nanotube devices can be made with a range of nanotube
diameters, offering different bandgaps.  We find the scattering
time to be approximately proportional to the diameter for all
charge densities and temperatures. Indeed, the combination
$\tau/d$ depends primarily on temperature and charge density and
only weakly on diameter as shown in Fig.~\ref{figtau}. (In the
zero density limit, there is an additional simplification
\cite{Perebeinos3}, $\tau /d\propto T^{-1}$.)

\begin{figure}[h!]
\includegraphics[height=2.8in]{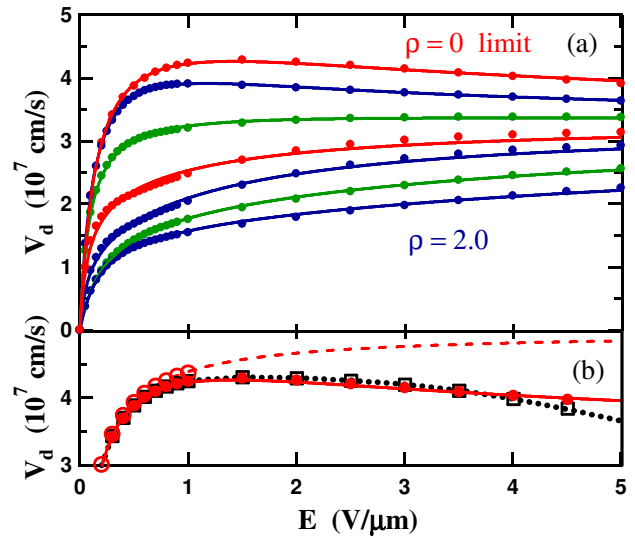}
\caption{\label{figvel} (a) Drift velocity (filled circles) at
$T=300$ K, as a function of electric field, in (19,0) tube
($d=$1.5 nm) at different charge densities. Curves show best fit
to Eq.~\protect{\ref{eqvel}}. Charge densities (in electrons/nm)
are, from top to bottom, $\rho=0$ limit (red), $\rho=0.25$ (blue),
$0.5$ (green), $0.75$ (red), $1.0$ (blue), $1.5$ (green), $2.0$
(blue). (b) The $\rho=0$ results from (a) are repeated here (red
solid curve, filled circles), and compared with different
simplified models: single band (black open squares); single
hyperbolic band (black dotted curve); and single parabolic band
(red open circles), extrapolated as red dashed curve using the
best fit to Eq.~(\protect{\ref{eqvel}}) , i.e. $V_{s1}=5.0 \ 10^7$
cm/s and $E_2=\infty$.}
\end{figure}

So far we have focused on the transport properties at low fields.
Transport at high fields is also important, and our results for
the dependence of carrier velocity on electric field are
summarized in Fig.~\ref{figvel}. For low carrier density, the
drift velocity has a maximum at a field of $E\sim$ 0.5-1.0
(V/$\mu$m), which is within the range of fields used in normal
device operation, and a negative differential mobility (NDM) at
higher fields. The NDM becomes weaker with increasing density, and
disappears entirely at a critical density $\rho_c$. We find that
$\rho_c$ is not a strong function of tube diameter $d$; for the
range of $d$ considered here, $\rho_c\sim$0.35-0.5 electrons/nm at
room temperature. Above $\rho_c$, the drift velocity $V_d$ is a
monotonic function of electric field, with no saturation behavior
up to the fields studied here \cite{disclaimer2}.

We can fit all the results in Fig.~\ref{figvel}a using the
following phenomenological formula:
\begin{eqnarray}
V_d=\frac{\mu_0E+V_{s2}\left(\left.{E}\right/{E_2}\right)^2}{1+\left.{\mu_0E}\right/{V_{s1}}+
\left(\left.{E}\right/{E_2}\right)^2} \label{eqvel}
\end{eqnarray}
where $\mu_0$ is the low-field mobility (from
Fig.~\ref{figdecom}), and $V_{s2}$ is the saturation velocity at
$\rho_c$. The value of $V_{s2}=3 \times 10^7$ (cm/s) is about
$1/3$ of the Fermi velocity for graphene and it is roughly
independent of tube diameter. The other two parameters, $V_{s1}$
and $E_2$, are fitted separately for each density.

A recent experiment \cite{ChenF} at high drain voltage in CN-FETs
(with $d\sim$ 2.0-2.4 nm) gave a saturation velocity of $2 \times
10^7$ cm/s. This is a factor of 2.5 smaller than the maximum
velocity predicted in the zero density limit
\cite{Perebeinos2,Pennington}. The electrostatic gate capacitance
per unit length was estimated to be $C_g=2\times10^{-11}$ F/m.
This gives a charge density $\rho=C_gV_g=1.1$ electrons/nm for the
gate voltage of 9 V \cite{ChenF}. This implies that the
experimental device was in the regime, where the carrier density
$\rho$ was well above critical density $\rho_c$. Therefore, the
zero density limit is not applicable, and the finite density
result obtained from Fig.~\ref{figvel} agrees much better with the
experiment.

Negative differential mobility has been predicted theoretically
\cite{Pennington,Perebeinos3,Verma} and negative differential
conductance has been observed experimentally \cite{Dai}. One
possible interpretation of this behavior \cite{Pennington,Verma}
is that the occupation of the second band (which has higher
effective mass) is responsible for the NDM. In agreement with
previous studies \cite{Pennington,Verma}, we find that at high
field a significant fraction of charge carriers occupy the second
band. However, we find that this alone does not explain the NDM.
In fact, if we artificially restrict the electrons to the first
band, the NDM becomes stronger rather than weaker
(Fig.~\ref{figvel}b).

To understand the origin of the NDM, we examine the effect of
changes in bandstructure. The tight-binding  bandstructure is well
described by a hyperbolic dispersion [Eq.~(\ref{eq1})]; and
Fig.~\ref{figvel}b shows that the hyperbolic bandstructure gives
results essentially identical to the real bandstructure (when only
the first band is included). However, when we replace the
realistic hyperbolic bandstructure with a parabolic band in
Fig.~\ref{figvel}b, the NDM disappears entirely. Therefore, we
conclude that the non-parabolicity of the first band is
responsible for the NDM in semiconducting carbon nanotubes.

In the NDM regime, the mobility is approximately constant over a
wide range of electric field, and equal to the most negative
value, i.e. the maximum NDM. Therefore, we can characterize the
NDM of a given tube by two parameters: the most-negative value
$\mu_n$; and the field at which the velocity reaches a maximum.

\begin{figure}[h!]
\includegraphics[height=3.4in]{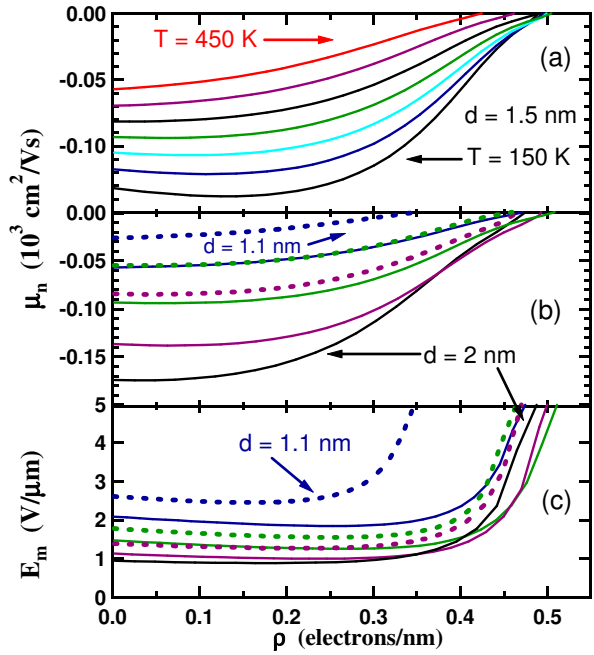}
\caption{\label{figndr}  (a) Maximum NDM, $\mu_n$, as a function
of charge density in (19,0) tube for different temperatures from
$T=450$ K (top) to $T=150$ K (bottom) with in steps of $50$ K. (b)
$\mu_n$ as a function of charge density in zig-zag tubes at
$T=300$ K. The mod$(n,3)=1$ tubes are shown by the solid curves
$n=16$ (blue), 19 (green), 22 (magenta), 25 (black), and
mod$(n,3)=-1$ tubes are by the dashed curves n=17 (blue), 20
(green), 23 (magenta). (c) The electric field corresponding to the
maximum drift velocity in the same tubes as in (b).}
\end{figure}

With these parameters, we can summarize the behavior over a range
of density,  different temperatures and tube diameters as shown in
Fig.~\ref{figndr}.  As the temperature is increased, the maximum
NDM decreases  approximately linearly with temperature.

The maximum NDM, shown in Fig.~\ref{figndr}b, increases with tube
diameter and has a clear family effect. The family of tubes with
mod$(n-m,3)=1$ have higher NDM than family of tubes with
mod$(n-m,3)=-1$. This can be understood by the fact that the ratio
$\Delta/m_0$ measures the non-parabolicity of the band, which
depends on the tube family. In the calculations for a single
hyperbolic band Eq.~(\ref{eq1}), we find that the NDM is a linear
function of $\Delta/m_0$, even though $\Delta$ and $m_0$ are
independent variables. If $\Delta$ and $m_0$ coincide with the
tight-binding values, the drift velocity is virtually the same as
in the  tight-binding band  calculations restricted to a single
band (as show in Fig.~\ref{figvel}b).

Finally, the velocity reaches  a maximum at a field $E_m$ shown in
Fig.~\ref{figndr}c, which decreases with the tube diameter. This
suggests that conditions for the NDM are easier to achieve in
larger diameter tubes. At the critical density $\rho_c$, the field
$E_m$ grows to infinity and  NDM disappears.

In conclusion, we calculated the transport properties of  CN-FETs
as a function of change density as controlled by the gate
potential. The low-field mobility dependence on the charge density
is mainly determined by the scattering at low temperatures, and by
both scattering and non-parabolicity of the first conduction band
at high temperatures. At high fields, the drift velocity shows a
negative differential mobility behavior which is large in large
diameter tubes and low temperatures. We showed that the
non-parabolicity of the first conduction band is responsible for
the NDM behavior.


\begin{thebibliography}{}
\bibitem{disclaimer} Electron-electron scattering is neglected here,
since it does not change the net momentum of the interacting
charge carriers.
\bibitem{disclaimer2}
At electric fields above a few V/$\mu$m, the charge carrier can be
accelerated to energies above the bandgap. This leads to impact
excitation (as evidenced by an increase in electroluminiecence
\protect{\cite{Jia}}). This additional decay channel is not
included in our description, and, therefore, we report the
calculations for the fields up to $5$ (V/$\mu$m) only.
\end{thebibliography}
\end{document}